\documentclass[%
 aip,
 amsmath,amssymb,
 reprint,%
]{revtex4-1}

\usepackage{graphicx}
\usepackage{dcolumn}
\usepackage{bm}

\usepackage[utf8]{inputenc}
\usepackage[T1]{fontenc}
\usepackage{mathptmx}
\usepackage{etoolbox}
\usepackage{color}

\makeatletter
\def\@email#1#2{%
 \endgroup
 \patchcmd{\titleblock@produce}
  {\frontmatter@RRAPformat}
  {\frontmatter@RRAPformat{\produce@RRAP{*#1\href{mailto:#2}{#2}}}\frontmatter@RRAPformat}
  {}{}
}%
\makeatother
\begin{document}

\preprint{AIP/123-QED}

\title[Synchronization of phase oscillators on complex hypergraphs]{Synchronization of phase oscillators on complex hypergraphs}
\author{Sabina Adhikari}%
\email{sabina.adhikari@colorado.edu}
\author{Juan G. Restrepo}
\affiliation{ 
Department of Applied Mathematics, University of Colorado Boulder, Boulder, CO 80309 
}%

\author{Per Sebastian Skardal}
\affiliation{%
Department of Mathematics, Trinity College, Hartford, CT 06106
}%

\date{\today}

\begin{abstract}
We study the effect of structured higher-order interactions on the collective behavior of coupled phase oscillators. By combining a hypergraph generative model with dimensionality reduction techniques, we obtain a reduced system of differential equations for the system's order parameters. We illustrate our framework with the example of a hypergraph with hyperedges of sizes 2 (links) and 3 (triangles). For this case, we obtain a set of 2 coupled nonlinear algebraic equations for the order parameters. For strong values of coupling via triangles, the system exhibits bistability and explosive synchronization transitions. We find conditions that lead to bistability in terms of hypergraph properties and validate our predictions with numerical simulations. Our results provide a general framework to study synchronization of phase oscillators in hypergraphs, and they can be extended to hypergraphs with hyperedges of arbitrary sizes, dynamic-structural correlations, and other features. 

\end{abstract}

\maketitle

\begin{quotation}

Synchronization of networks of coupled oscillators is one of the most iconic problems in complex systems, with applications in biology\cite{petri2014homological,kitzbichler2009broadband}, physics \cite{zhu2015synchronization} and engineering \cite{rohden2012self,fujino1993synchronization,strogatz2005crowd}. Usually, coupling between oscillators is assumed to be mediated by pair interactions. Recently, motivated by applications in physics\cite{ashwin, pazo} and biology\cite{petri2014homological, Brain-simplex}, there has been much interest in studying the effect of higher-order interactions, i.e., simultaneous interactions between multiple oscillators, on synchronization patterns\cite{skardal-arenas1, app-simplex1,skardal-arenas,millan2020explosive,ghorbanchian2021higher, calmon2021topological}. In this paper, we study synchronization of coupled phase oscillators on complex hypergraphs. We use a hypergraph generative model and develop a mean-field analysis using dimensionality reduction techniques to obtain low-dimensional descriptions of synchronization in terms of hypergraph structural parameters. We find conditions on the hypergraph that result in bistability. Our results provide a general and flexible framework to study synchronization on hypergraphs.

\end{quotation}

\section{\label{sec:level1}Introduction\protect\\}

Synchronization processes are present in many applications\cite{Opisov_book, arenas2008synchronization}. Some common examples include collections of flashing fireflies \cite{Fireflies,sarfati2020spatio}, crickets chirping in unison \cite{Opisov_book}, neuronal networks \cite{neurons1}, cortical brain rhythms \cite{petri2014homological}, and power grid dynamics \cite{Powergrid,Powergrid1}. A paradigmatic model for synchronization is the Kuramoto model of phase oscillators \cite{kuramoto1975self,acebron2005kuramoto}, in which synchronization is mediated by pairwise interactions between oscillators. The Kuramoto model on complex networks has many applications and is one of the central models in complex science \cite{rodrigues2016kuramoto, PhysRevLett.106.128701, laing2009chimera}.
Recently, with motivation from fundamental principles \cite{ashwin, pazo} and applications to neuroscience \cite{petri2014homological, Brain-simplex}, there has been much attention devoted to synchronization in networks with higher-order interactions, i.e., simultaneous interactions between multiple nodes. Higher-order interactions in coupled phase oscillator systems result in interesting phenomena like abrupt switching between incoherent and synchronized states, hysteresis, and bistability \cite{skardal-arenas1, skardal-arenas}. So far, most of the analytical results have been obtained for the all-to-all coupling case, and there is not a clear way to predict the effect of complex interaction structure on these phenomena. 
In this paper we study synchronization of phase oscillators on complex hypergraphs, i.e., networks with higher-order interactions and non-trivial connectivity. To do so, we restrict our attention to a specific but flexible hypergraph generative model that allows us to generate and study hypergraphs with tunable characteristics. By using this generative model in combination with the Ott-Antonsen ansatz \cite{Ott-Antonson}, we are able to obtain low-dimensional descriptions of the system's order parameters in terms of the hypergraph's structural properties. We illustrate our approach with two examples of a hypergraph with interactions of sizes 2 (links) and 3 (triangles): a random hypergraph and a hypergraph constructed in such a way that the numbers of links and triangles at each node are correlated. We derive analytical conditions on the properties of these hypergraphs that result in synchronization, incoherence, or bistable behavior and validate our results with numerical simulations.

The paper is organized as follows. In section \ref{Model}, we present our hypergraph generative model and the Kuramoto model on hypergraphs. In section \ref{Meanfield}, we use the Ott-Antonsen ansatz and a mean-field approximation to obtain low dimensional descriptions for the  local and global order parameters. In section \ref{Examples}, we demonstrate our framework on two example hypergraphs.  In section \ref{Discussion},
we discuss our results and their limitations. 

\section{Model} \label{Model}

In this section we introduce the hypergraph generative model and the Kuramoto model on hypergraphs.

\subsection{Hypergraph model}\label{hypergraphmodel}

A hypergraph is a pair of nodes and hyperedges $(V,E)$, where $V$ is the set of nodes labeled $n = 1,2\dots, N$, and the set of hyperedges $E$ is a set of subsets of $V$. The $m$'th order degree of a node $n$ is given by $k_{n}^{(m)}$, which gives the number of hyperedges with size $m$ that node $n$ is a part of. The {\it hyperdegree} of node $n$ is given by ${\bf k}_n = \{k_n^{(1)}, k_n^{(2)}, ..., k_n^{(M)}\}$, where $M$ is the largest hyperedge size.  For simplicity, we refer to hyperedges of sizes $2$ and $3$ as {\it links} and {\it triangles} respectively. We denote by $N({\textbf{k}})$ the number of nodes with hyperdegree ${\bf k}$, and define the hyperdegree distribution as $P({\bf k}) = N({\bf k})/N$.

We will consider synchronization on a class of hypergraphs produced by the following generative model. For a given set of nodes $n = 1,2,\dots, N$ and a specified vector of target hyperdegrees $[{\bf k}_1,{\bf k}_2,\dots,{\bf k}_N]$, the hyperedge $\{i_1,i_2,\dots,i_m\}$ is created with probability $a^{(m)}({\bf k}_{i_1},{\bf k}_{i_2},\dots,{\bf k}_{i_m})$. By counting the expected number of hyperedges of size $m$ in two different ways, one finds that the functions $a^{(m)}$ should be normalized such that 
\begin{align}
&\frac{1}{m!}\sum_{{\bf k}_1,\dots,{\bf k}_m}N({\bf k}_1)\cdots N({\bf k}_m)a^{(m)}({\bf k}_{i_1},{\bf k}_{i_2},\dots,{\bf k}_{i_m})\nonumber \\
= &\frac{1}{m} \sum_{\bf k} N({\bf k}) k^{(m)}.
\label{eq:hyperedge}
\end{align}
This model is a natural extension of latent feature models \cite{miller2009nonparametric} to hypergraphs, and allows us to generate hypergraphs with heterogeneous and correlated hyperdegree distributions \cite{Nick}. The model can be easily extended to the case where hyperedges connect preferentially nodes with certain attribute variables such as nodal community index, oscillator frequency, or other dynamical parameters.  On the other hand, the generative model is not able to capture features beyond the preference for hyperedges to connect certain types of nodes. An important class of hypergraphs that is not captured by this generative model is that of simplicial complexes, where triangles only connect triads of nodes that form a clique with pairwise connections (for simplicial complex generative models, see for example Refs.\cite{bianconi2016network,courtney2016generalized,kovalenko2021growing,Higher-order-networks}).

Our subsequent results will apply to the ``expected'' network generated from this generative model. Such an approach is similar to the analysis of network processes based on the configuration model (e.g., Ref.~\cite{pastor2015epidemic}) or the {\it annealed network approximation} \cite{dorogovtsev2008critical, boguna2009langevin, poux2020influential}. A similar approach has been successfully applied to the Kuramoto model on pairwise networks \cite{restrepo2014mean}. The limitations of this approach are discussed in Sec.~\ref{Discussion}.

\subsection{Higher-order Kuramoto Model}

The Kuramoto model of phase synchronization can be generalized to account for higher order interactions in different ways. In Refs.~\cite{millan2020explosive,ghorbanchian2021higher, calmon2021topological}, the synchronization of phases defined on the faces of a simplicial complex is studied. Here, following \cite{skardal-arenas1, skardal-arenas}, we will instead consider synchronization mediated by the simultaneous, nonlinear interaction of all the phases belonging to the edges of a hypergraph. In this context, the Kuramoto model for the phases $\theta_n$ of nodes $n = 1,2,\cdots,N$ on a hypergraph $H$ can be generalized to 
\begin{align}
\frac{d\theta_n}{dt} = \omega_n + \sum_{n\in e\in E} K_e \sum_{\mathbb{P}} \sin({\bf c}^T_e \mathbb{P} {\vec \theta}_e), \label{eq:kurahyper}
\end{align}
where the coupling term sums over all edges $e\in E$ containing node $n$, $K_e$ is the coupling to edge $e$, ${\vec \theta}_e$ is the vector of phases of oscillators in edge $e$ with $\theta_n$ placed in the last component, and $\mathbb{P}$ is a permutation of the remaining components. Adding over all permutations $\mathbb{P}$ ensures symmetric coupling from all the other nodes in the hyperedge. The integer-valued vector ${\bf c}_e$ determines how the phases are combined inside the sine function and satisfies ${\bf c}_e^T {\bf 1} = 0$, where ${\bf 1}$ is a vector of ones. In the case of the pairwise all-to-all Kuramoto model, for example,  $K_e = K/N$ and ${\bf c}_e = [1,-1]^T$. 

Here we will study the case where there are hyperedges only of sizes 2 (links) and 3 (triangles), $K_e = K_2$, ${\bf c}_e = [1,-1]^T$ for links, and $K_e = K_3$, ${\bf c}_e = [2,-1,-1]^T$ for triangles. With these choices, Eq.~(\ref{eq:kurahyper}) can be rewritten as
\begin{eqnarray}
\frac{d\theta_n}{dt}  &=& \omega_n + K_2 \sum_{m=1}^N A_{nm} \sin{(\theta_m - \theta_n)} + \nonumber \\
    & & K_3 \sum_{j,m} B_{njm} \sin{(2\theta_j - \theta_m - \theta_n)},
    \label{eq1}
\end{eqnarray}
where we assume that the hypergraph is described by symmetric tensors with entries $A_{nm}$ and $B_{njm}$, where $A_{nm} = 1$ ($0$) if nodes $n, m$ are connected (not connected) by a link, and $B_{njm} =1 (B_{njm} = 0)$ if nodes $n,j,m$ are connected (not connected) by a triangle. However, the techniques that we present can be applied to the general case (\ref{eq:kurahyper}) as long as the hypergraph is generated (or can be approximated) with a generative model like the one discussed in Sec.~\ref{hypergraphmodel}.

Higher order interactions of the form Eq.~(\ref{eq1}) can arise when a phase oscillator model is derived from an expansion beyond first order of the complex Ginzburg-Landau equation (e.g., see Refs. \cite{ashwin, pazo}). The diffusive-type coupling case where ${\bf c}_e = [1,1,-2]^T$ for triangles has been studied for the all-to-all case in Ref.~\cite{skardal-arenas}, so here we focus for simplicity on the form of the interactions in Eq.~(\ref{eq1}).

\section{Dimensionality Reduction} \label{Meanfield}
In this section we use the Ott-Antonsen Ansatz{\cite{Ott-Antonson}} to derive a low-dimensional description of the dynamics and use it to find semi-analytical expressions for the order parameters. In order to accomplish this, we use a generalization of the Ansatz in which oscillators are divided into subgroups of oscillators with the same hyperdegree, with oscillators in each subgroup assumed to be statistically equivalent \cite{Kuramoto-bick, Kuramoto-pikovsky, restrepo2014mean}. Furthermore, we neglect pair correlations among oscillators connected in triangles. These approximations are presented and discussed below. Using this procedure, we obtain a low dimensional description in terms of the functions determining the probabilities of connection between the different subgroups, i.e., the functions $a^{(m)}$. This low dimensional description allows us to find conditions for synchronization and for the appearance of bistability of the synchronized and incoherent states. 

Defining the  local order parameters
 \begin{eqnarray}
 R^{(1)}_n &=&\sum_{m} A_{nm} e^{i \theta_m}, \hspace{0.5cm}
 R^{(2)}_n = \sum_{m,j} B_{njm} e^{2 i \theta_j} e^{-i \theta_m},
 \label{eq2}
 \end{eqnarray}
we can rewrite Eq.~(\ref{eq1}) as
\begin{eqnarray}
\frac{d{\theta_n}}{dt} = \omega_n + \text{Im}(H_ne^{-i\theta_n}),
\label{eq3}
\end{eqnarray}
where we defined $H_n = K_2 R_n^{(1)} + K_3 R_n^{(2)}$.

Following \cite{restrepo2014mean}, now we assume, based on the construction of the hypergraph from the generative model in Sec.~\ref{hypergraphmodel}, that nodes with the same hyperdegree ${\bf k}$ are statistically equivalent, and make the identification
\begin{equation}
    \begin{split}
    R^{(1)}_n \to R^{(1)} ({\bf k}_n,t),\\
    R^{(2)}_n \to R^{(2)} ({\bf k}_n,t).
    \end{split}
     \label{eq4}
\end{equation}
Moving to the continuum description in the limit as $N \to \infty$, we define $f(\theta, \omega, {\bf k}, t)$ to be the density of oscillators with phase $\theta$, natural frequency $\omega$, and hyperdegree $\bf k$ at time $t$. Thus, we divide the population of oscillators into subpopulations characterized by their hyperdegree, which acts as a population parameter as in Refs.~\cite{Kuramoto-bick, Kuramoto-pikovsky, restrepo2014mean}. In a mean-field approximation, the global order parameter $R^{(1)}{(\bf{k})}$ can be written in terms of the connection probabilities $a^{(2)}({\bf k},{\bf k}')$ and $a^{(3)}({\bf k},{\bf k}', {\bf k}'')$ introduced in Sec.~\ref{hypergraphmodel} as
\begin{eqnarray}
R^{(1)}(\bf{k}) &=& \sum_{\bf{k'}} {N(\bf k' )} a^{(2)}({\bf k},{\bf k'}) \iint f(\theta', \omega', {\bf k'}, t) e^{i \theta'} d\theta' d\omega'. \nonumber\\
\label{eq05}
\end{eqnarray}
Similarly, the global order parameter $R^{(2)}{(\bf{k})}$ can be written in terms of the joint density of two oscillators, $f_2(\theta, \omega, \theta', \omega',{\bf k}, {\bf k'}, t)$ as
\begin{widetext}
\begin{eqnarray}
R^{(2)}(\bf{k}) &=& {\sum_{\bf{k', k''}}} N({\bf k'}) N({\bf k''}) {a^{(3)}}({\bf k}, {\bf k'}, {\bf k''}) \iiiint f_2(\theta', \omega', \theta'', \omega'',{\bf k'}, {\bf k''}, t) e^{2i \theta'} e^{-i \theta''} d\theta' d\omega' d\theta'' d\omega''
\label{eq06}\\
&\approx& \sum_{{\bf k'}, {\bf k''}} N({\bf k'}) N({\bf k''}) {a^{(3)}} ({\bf k}, {\bf k'}, {\bf k''}) \iint f(\theta', \omega', {\bf k'}, t) e^{2 i \theta'} d\omega' d\theta' \iint f(\theta'', \omega'', {\bf k''}, t) e^{- i \theta''} d\omega'' d\theta'',
\nonumber
\end{eqnarray}
\end{widetext}
where, to make further progress, we have neglected pair correlations and assumed that the joint density can be written as 
\begin{align}
f_2(\theta', \omega', \theta'', \omega'',{\bf{k'}, \bf{k''}}, t) = f(\theta', \omega', {\bf{k'}}, t)f(\theta'', \omega', {\bf{k''}}, t). \label{approximation}
\end{align}
We offer the following heuristic arguments to support this assumption: first, in the limits of total incoherence and total synchronization Eq.~(\ref{approximation}) is exact. Second, when each oscillator is connected to many others, the correlations between any specific pair of oscillators should be small. Thus, we anticipate that this approximation will be a good one either close to total synchrony or incoherence, or for dense hypergraphs. This approximation is discussed further in the Discussion. For the regular Kuramoto model the effect of including pair (and higher) correlations has been studied in Ref.~\cite{hildebrand2007kinetic}.

Due to conservation of oscillators, the evolution of $f$ is governed by the continuity equation
\begin{eqnarray}
\frac{\partial f}{\partial t} + \frac{\partial}{\partial \theta} \bigg\{(\omega + \text{Im}[H e^{-i \theta_n}])f\bigg\} &=& 0.
\label{eq07}
\end{eqnarray}
To reduce the dimensionality of this system, we write $f$ as a Fourier series,
\begin{eqnarray}
 f &=& \frac{g(\omega)}{2 \pi} \left[ 1 + \sum_{n=1}^{\infty} b_n(\omega,{\bf k}, t) e^{-i n \theta} + \text{c.c.}\right],
 \label{eq08}
\end{eqnarray}
where c.c. denotes complex conjugate, and use the Ott-Antonsen ansatz{\cite{Ott-Antonson}} $b_n(\omega, {\bf k}, t) = (b(\omega, {\bf k}, t))^n$. Substituting this ansatz in Eq.~(\ref{eq07}), one finds that the continuity equation is satisfied if $b(\omega, {\bf k},t)$ satisfies the ODE
\begin{eqnarray}
\frac{d b}{d t} + i \omega b -\frac{1}{2} [H-H^*b^2] &=& 0.
\label{eq11}
\end{eqnarray}
Substituting Eq.~(\ref{eq08}) in Eqs.~(\ref{eq05})
and (\ref{eq06}), we obtain
\begin{eqnarray}
R^{(1)}({\bf k}) &=& \sum_{\bf k'} N({\bf k'}) a^{(2)} ({\bf k}, {\bf k'}) \int g(\omega') b(\omega',{\bf k'},t) d\omega', \label{eq09}\\
R^{(2)}({\bf k}) &=&  \sum_{{\bf k', k''}} N({\bf k'}) N({\bf k''}) a^{(3)} ({\bf k}, {\bf k'}, {\bf k''}) \int g(\omega') b^2(\omega',{\bf k'},t) d\omega'\nonumber\\
& &\int g(\omega'') b^*(\omega'',{\bf k''},t) d\omega''.
\label{eq10}
\end{eqnarray}

Assuming a Lorentzian distribution of frequencies, $g(\omega) = \Delta/(\pi[\Delta^2 + (\omega-\omega_0)^2])$ and using contour integration to evaluate the integrals in Eqs. (\ref{eq09}) and (\ref{eq10}), we get
\begin{eqnarray}
 R^{(1)}({\bf k}) &=& \sum_{{\bf k'}} N({\bf k'}) a^{(2)} ({\bf k}, {\bf k'}) b(\omega_0 - i\Delta, {\bf k'}, t),
 \label{eq12}
\end{eqnarray}
\begin{eqnarray}
 R^{(2)}({\bf k}) &=& \sum_{\bf {k',k''}} N({\bf k'}) N({\bf k''}) a^{(3)}({\bf k}, {\bf k'}, {\bf k''})\times \nonumber\\
 & &b^2(\omega_0 - i\Delta, {\bf k'}, t) b^*(\omega_0 - i\Delta, {\bf k''}, t).
 \label{eq13}
\end{eqnarray}
Inserting these in Eq.~(\ref{eq11}) and letting $\omega = \omega_0 - i \Delta$ and $b({\bf k},t) = b(\omega_0 - i \Delta, {\bf k}, t)$ we get
\begin{widetext}
\begin{eqnarray}
0 &=& \frac{d b({\bf k})}{d t} + i(\omega_0 - i \Delta)b({\bf k}) - \frac{K_2}{2} \sum_{{\bf k'}} N({\bf k'}) a^{(2)}({\bf k}, {\bf k'}) [b({\bf k'}) - b({\bf k'})^* b^2({\bf k})] \nonumber\\
&&
- \frac{K_3}{2} \sum_{{\bf k'}, {\bf k''}} N({\bf k'})N({\bf k''})  a^{(3)}({\bf k}, {\bf k'}, {\bf k''})[b^2({\bf k'}) b({\bf k''})^* - b^2({\bf k'})^* b({\bf k''})b^2({\bf k})].
\label{eq14}
\end{eqnarray}
\end{widetext}
Eq.~(\ref{eq14}) provides a low-dimensional description of the dynamics in terms of the hypergraph generative functions $a^{(2)}$ and $a^{(3)}$. While the number of variables $b({\bf k})$ might still be large, Eq. (\ref{eq14}) allows us to study the bifurcations and fixed points of the system. For this, it is useful to define the global order parameters
\begin{eqnarray}
R^{(1)} = \frac{1}{N\langle {\bf k}^{(1)}\rangle}\sum_{n=1}^N R_n^{(1)}, \quad R^{(2)} = \frac{1}{2 N\langle {\bf k}^{(2)}\rangle}\sum_{n=1}^N R_n^{(2)},
\label{eq34}
\end{eqnarray}
which can be written in terms of $b({\bf k}, t)$ as
\begin{eqnarray}
R^{(1)}(t) &=& \frac{1}{N \langle {\bf k}^{(1)} \rangle} \sum_{{\bf k}, {\bf k'}} N({\bf k}) N ({\bf k'}) a^{(2)} ({\bf k}, {\bf k'}) b({\bf k'},t),
\label{eq35}
\end{eqnarray}
\begin{eqnarray}
R^{(2)}(t) &=& \frac{1}{2 N \langle {\bf k}^{(2)} \rangle} \sum_{{\bf k}, {\bf k'}, {\bf k''}} N({\bf k}) N ({\bf k'}) N ({\bf k''}) \nonumber \\
& & a^{(3)} ({\bf k}, {\bf k'}, {\bf k''}) b^2({\bf k'},t)b^*({\bf k''},t).\label{R2}
\label{eq36}
\end{eqnarray}
The factor of $2$ in the definition of $R^{(2)}$ accounts for the fact that each triangle is counted twice in the calculation of $R_n^{(2)}$; note that in the case of complete synchronization, $b = 1$, the normalization (\ref{eq:hyperedge}) ensures $R^{(2)} = 1$. 

In the following, we will demonstrate the application of this formalism to selected examples. 

\section{Examples} \label{Examples}
In this section, we apply our theory to two examples: a random hypergraph analogous to an Erd\"os-R\'enyi network, and a hypergraph where the triangle and link degrees are correlated.

\subsection{Random hypergraph}

\begin{figure}[t]
\includegraphics[clip,width=0.8\columnwidth]{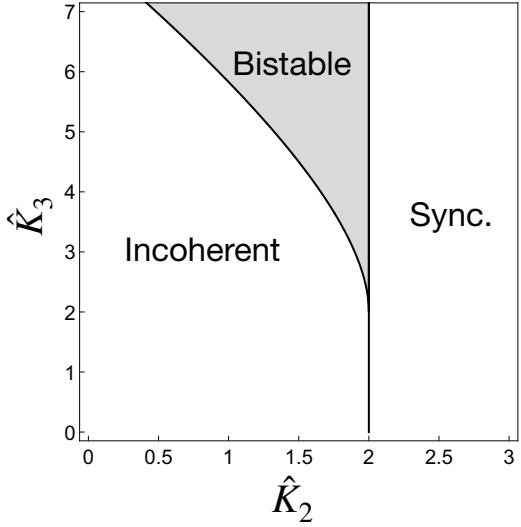}
\caption{\label{fig:phase} Phase diagram for a random hypergraph as a function of the parameters $\hat K_2 = \langle k \rangle K_2$ and $\hat K_3 = 2\langle q \rangle K_3$. The bistable region is separated from the incoherent region by a saddle-node bifurcation, and from the synchronized region by a subcritical pitchfork bifurcation. The incoherent and synchronized regions are separated by a supercritical pitchfork bifurcation.}
\end{figure}

We start by considering the hypergraph analog of an Erd\"os-R\'enyi network, i.e., a hypergraph where a link connects every pair of nodes with probability $p_2$ and a triangle connects every triad of nodes with probability $p_3$. Synchronization on this hypergraph was studied numerically in Ref.\cite{skardal-arenas1}. In terms of the average numbers of links and triangled per node, $\langle k \rangle$ and $\langle q \rangle$, using Eq.~(\ref{eq:hyperedge}), we obtain
\begin{align}
p_2 &= a^{(2)}({\bf k},{\bf k}') = \frac{\langle k \rangle}{N}\\
p_3 &= a^{(3)}({\bf k},{\bf k}',{\bf k}'') = \frac{2 \langle q \rangle}{N^2}, 
\end{align}
where we assumed $N \gg 1$. Inserting these in Eq.~(\ref{eq14}), we find that all $b{({\bf k})}$  satisfy the same equation, 
\begin{align}
0 = \frac{d b({\bf k})}{dt} &+ \Delta b({\bf k}) + i \omega_0 b{\bf(k)}  -\frac{K_2 \langle k \rangle}{2}[V_1 - b^2{\bf(k)} V_1^* ]\nonumber \\
&- K_3 \langle q \rangle[V_1^* V_2 - b^2{\bf(k)} V_2^* V_1],\label{eqbs}
\end{align}
where
\begin{eqnarray}
V_1 &= \frac{1}{N}\sum_{{\bf k}'} N({\bf k}') b({\bf k}'), \label{eqv1}\\
V_2 &= \frac{1}{N}\sum_{{\bf k}'} N({\bf k}') b({\bf k}')^2. \label{eqv2}
\end{eqnarray}
Since all $b({\bf k})$ approach the same attractors, we look for stationary rotating solutions of the form $b({\bf k}) = b e^{i \Omega t}$, $V_1(t) = V_1 e^{i \Omega t}$, $V_2(t) = V_2 e^{2i \Omega t}$. Each $b({\bf k})$ is assumed to have the same complex phase dictated by the fourth and fifth terms of Eq.~(\ref{eqbs}). After separating real and imaginary parts, we find that $\Omega = -\omega_0$ and that $b$ satisfies $b = 0$ or (note that $V_1 = b$, $V_2 = b^2$)
\begin{align}
0 = b^2(K_2 \langle k \rangle + 2 K_3 \langle q \rangle b^2) + 2  - (K_2 \langle k \rangle  + 2 K_3 \langle q \rangle b^2).\nonumber
\end{align}
Solving for $b$ and noting that, from Eq.~(\ref{eq35}), $|R^{(1)}| = b$, we get 
\begin{align}
|R^{(1)}| = \sqrt{\frac{\hat K_3  -\hat K_2 \pm \sqrt{\left(\hat K_2+ \hat K_3 \right)^2-8 \hat K_3 }}{2 \hat K_3}},\label{eqr1}
\end{align}
where $\hat K_2 = \langle k \rangle K_2$ and $\hat K_3 = 2 \langle q \rangle K_3$.
This generalizes the all-to-all result [Eq.~(5) in Ref.\cite{skardal-arenas1}] to random hypergraphs by properly rescaling the dyadic and triadic coupling strengths. (The factor of $2$ in $\hat K_3$ can be understood in the context of the normalization used in Ref.\cite{skardal-arenas1}  by noting that in the all-to-all case the mean triangle degree is $\langle q \rangle \approx N^2 /2$.)  Depending on the values of $\hat K_2$ and $\hat K_3$, Eq.~(\ref{eqr1}) can have zero, one, or two real solutions. As noted in Ref.\cite{skardal-arenas1}, for $\hat K_3 < 2$ the system undergoes a supercritical pitchfork bifurcation from incoherence ($R^{(1)}$ = 0) to synchronization ($|R^{(1)}|$ > 0) at $\hat K_2 = 2$. For $\hat K_3 > 2$ the system is incoherent for $\hat K_2 < 2 \sqrt{2} \sqrt{\hat K_3}-\hat K_3$. At $\hat K_2 = 2 \sqrt{2} \sqrt{\hat K_3}- \hat K_3$ there is a saddle-node bifurcation where a pair of stable and unstable synchronized solutions appear. At $\hat K_2 = 2$, the unstable solution disappears in a subcritical pitchfork bifurcation. The phase diagram, mirroring that for the all-to-all case in Ref.\cite{skardal-arenas1}, is shown in Fig.~\ref{fig:phase}.

\subsection{Correlated links and triangles}

Now we move to an example where the structure of links is correlated with the structure of triangles. We assume that a prescribed degree sequence is given, $\{k_1, k_2, \cdots, k_N\}$, and links are created between nodes according to the Chung-Lu model~\cite{courtney2016generalized}, so that
\begin{eqnarray}
    a^{(2)} (k,k') &=& \frac{k k'}{N \langle k \rangle}.
    \label{eq37}
\end{eqnarray}
Following Ref.~\cite{Nick}, we consider a model where
 the probability that a triangle connects nodes with degrees $k, k'$ and $k''$ is given by
\begin{eqnarray}
a^{(3)}(k, k', k'') = \frac{2 k k' k''}{(N \langle k \rangle)^2}.
\label{eq40}
\end{eqnarray}
The normalization is chosen using Eq.~(\ref{eq:hyperedge}) so that $\langle k^{(2)}\rangle = \langle k^{(3)} \rangle = \langle k \rangle$.
\begin{figure}[t]
\includegraphics[clip,width=\columnwidth]{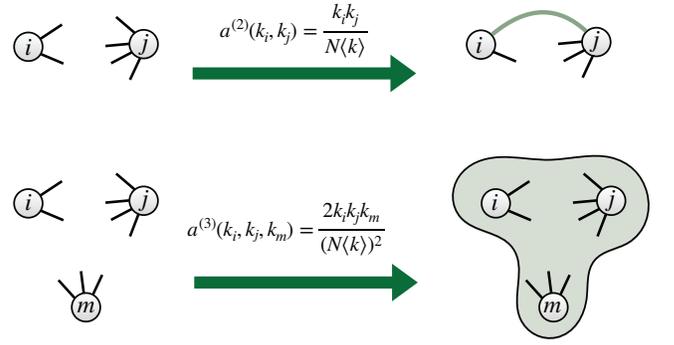}
\caption{\label{fig2} Schematic representation of the generative model for the correlated links and triangles hypergraph. (Top) Given a pair  of nodes with degrees $k_i, k_j$, a link is created with probability $a^{(2)} (k,k') = k k'/(N \langle k \rangle)$ (Ref.~\cite{courtney2016generalized}). (Bottom) Given three nodes with degrees $k_i, k_j,k_m$, a  triangle joining them is created with probability $a^{(3)}(k, k', k'') = 2 k k' k''/(N \langle k \rangle)^2$ (Ref.~\cite{Nick}).}
\end{figure}
By construction, the expected degrees $k^{(2)}_n$ and $k^{(3)}_n$ for a given node $n$ coincide, so we call this model {\it correlated links and triangles} \cite{Nick}. The model is illustrated in Fig.~\ref{fig2}. Since a node is only characterized by a single degree $k$, in the rest of this section we index all quantities by $k$ only, i.e., we write $b_k$ instead of b({\bf k}).  Using the forms for $a^{(2)}$ and $a^{(3)}$ above, Eq.~(\ref{eq14}) becomes
\begin{eqnarray}
0 &=&\frac{d b_k}{d t} + \Delta b_k + i \omega_0 b_k -\frac{K_2}{2} \sum_{k'} N(k') \frac{k k'}{N \langle k \rangle} [b_{k'} - b_{k'}^* b_{k}^2] \nonumber\\
& &
    - \frac{K_3}{2} \sum_{k', k''} N(k') N(k'') \frac{2 k k' k''}{(N \langle k \rangle)^2} [b_{k'}^2 b_{k''}^* - {b_{k'}^{2*}} b_{k''} b_{k}^2].
    \label{eq15}
\end{eqnarray}
Defining 
\begin{eqnarray}
U_1 &=& \sum_{k'} \frac{k' N(k') b_{k'}}{N \langle k \rangle}, \label{eq38}\\
U_2 &=& \sum_{k'} \frac{k' N(k') b_{k'}^2}{N \langle k \rangle}, \label{eq39}
\end{eqnarray}
Eq.~(\ref{eq15}) can be rewritten as
\begin{eqnarray}
0 &=& \frac{d b_k}{d t} + \Delta b_k + i \omega_0 b_k - \frac{K_2}{2} k U_1 + \nonumber \\ 
& &\frac{K_2}{2} k b_k^2 U_1^* - K_3 k U_2 U_1^* + K_3 k b_k^2 U_2^* U_1, \label{eq16}
\end{eqnarray}
where $\Delta$ comes from the Lorentzian distribution of frequencies ($g(\omega) = \Delta/(\pi[\Delta^2 + (\omega - \omega_0)^2])$). As compared to Eqs.~(\ref{eq09}) and (\ref{eq10}), the ODE on Eq.~(\ref{eq16}) is much simplified. We have reduced Eqs.~(\ref{eq05}) and (\ref{eq06}) to a closed set of ODEs in terms of variables $b_k(t)$ coupled to two global variables $U_1$ and $U_2$.
Now, seeking a stationary rotating solution, we let $b_k(t) = b_k e^{i \Omega t}$, $U_1(t) = U_1 e^{i \Omega t}$ and $U_2(t) = U_2 e^{i \Omega t}$. Then Eq.~(\ref{eq16}) becomes
\begin{eqnarray}
0 &=& i \Omega b_k e^{i \Omega t} + \Delta b_k e^{i \Omega t} + i \omega_0 b_k e^{i \Omega t} - \frac{K_2}{2} k U_1 e^{i \Omega t} \nonumber \\
&& + \frac{K_2 }{2} k b_k^2 U_1 e^{i \Omega t} - K_3 k U_2 U_1 e^{i \Omega t} + \nonumber\\
&& K_3 k b_k^2 U_2 U_1 e^{i \Omega t}.
 \label{eq18}
\end{eqnarray}
The imaginary part of Eq.~(\ref{eq18}) gives $\Omega = -\omega_0$, and the real part  simplifies to:
\begin{eqnarray}
0 &=& \Delta b_k - \frac{K_2}{2} k U_1 + \frac{K_2}{2} k b_k^2 U_1 - \nonumber\\
& & K_3 k U_2 U_1 + K_3 k b_k^2 U_2 U_1.
\label{eq19}
\end{eqnarray}
Solving for $b_k$, we get
\begin{eqnarray}
b_k(U_1,U_2) &=& \frac{-\Delta + \sqrt{\Delta^2 + \left[K_2kU_1 + 2K_3kU_2U_1 \right]^2}}{K_2kU_1 + 2K_3kU_2U_1},
\label{eq20}
\end{eqnarray}
where we chose the solution that satisfies $b_k \to 0$ when $K_3 = 0$, $K_2 \to 0$.
Inserting this expression into the definition of $U_1$ and $U_2$, we find the self-consistent equations
\begin{eqnarray}
U_1 &=& \frac{1}{N \langle k \rangle} \sum_{k} N(k) k b_k(U_1,U_2),
\label{eq21}\\
U_2 &=& \frac{1}{N \langle k \rangle} \sum_{k} N(k) k b_k^2(U_1,U_2).
\label{eq22}
\end{eqnarray}
Using Eqs.~(\ref{eq37}) and (\ref{eq40}) in Eqs.~(\ref{eq34}) and (\ref{eq35}), we find that the order parameters $R^{(1)}$ and $R^{(2)}$ can be expressed in terms of $U_1$ and $U_2$ as
\begin{eqnarray}
R^{(1)} &=& U_1,
\label{eq41}\\
R^{(2)} &=& U_2U_1^* \label{eq42}.
\end{eqnarray}
Note that setting $K_3=0$ in Eq.~(\ref{eq20}), one recovers after some manipulation the degree-based mean-field approximation for the network Kuramoto model [i.e., Eq.~(25) in Ref.\cite{Restrepo} or Eq.~(13) in Ref.\cite{Ichinomiya}].

With the self-consistent equations (\ref{eq20})-(\ref{eq22}), we now proceed to determine the nature of the bifurcation from incoherence ($U_1,U_2 = 0$) to synchronization ($U_1, U_2 > 0$) with a perturbative approach. Expanding (\ref{eq20}) for small $U_1$, $U_2$ we obtain up to cubic order in $U_1$ (note that $U_2 \sim U_1^2$)
\begin{align}
U_1 &= \frac{\langle k^2 \rangle}{2\langle k \rangle}K_2 U_1 + \frac{\langle k^2 \rangle}{\langle k \rangle} K_3 U_1 U_2 - \frac{\langle k^4 \rangle}{8\langle k \rangle} K_2^3 U_1^3, \\
U_2 &= \frac{\langle k^3 \rangle}{4 \langle k \rangle}K_2^3 U_1^2.\label{eq:expa}
\end{align}
Letting $U_1 \to 0^+$ to find the onset of synchronization, the leading order terms give the critical coupling strength
\begin{align}
K_2 = K_2^c &= \frac{2 \langle k \rangle}{\langle k^2 \rangle}.
\end{align}
Next, solving for $U_1$ we find, after canceling the incoherent solution, that close to the transition $U_1$ satisfies
\begin{align}
a U_1^2 = \frac{K_2}{K_2^c} -1,
\end{align} 
where 
\begin{align}
a = \left(\frac{\langle k^4 \rangle}{8 \langle k \rangle}K_2^3 - \frac{\langle k^2 \rangle \langle k^3 \rangle}{4 \langle k \rangle^2} K_2^2 K_3\right).\label{eq:a}
\end{align}
Thus, a bifurcation occurs at $K_2 = K_2^c$, which is independent of $K_3$ and equal to the critical constant for the network Kuramoto model in the mean-field approximation \cite{Ichinomiya, Restrepo}. The bifurcation is supercritical for $a > 0$ and subcritical for $a < 0$. Evaluating Eq.~(\ref{eq:a}) at $K_2 = K_2^c$, we find that the transition is subcritical (and therefore explosive and with hysteretic behavior) for 
\begin{equation}
K_3 > K_3^c = \frac{ \langle k^4 \rangle \langle k \rangle^2}{\langle k^2 \rangle^2 \langle k^3 \rangle} = \frac{ \langle k^4 \rangle \langle k \rangle}{2\langle k^2 \rangle \langle k^3 \rangle} K_2^c . \label{eq32}
\end{equation}

For regular networks with $k_i = k$, there is bistability for $K_3 > K_2^c/2 = 1/\langle k \rangle$. For networks with a diverging fourth moment [such as networks with a power-law degree distribution with exponent $\gamma \in (4,5)$ in the limit $N \to \infty$], $K_3^c$ diverges and there is no bistability.


Now we validate our theoretical results with numerical simulations. First, we generate a sequence of $N = 5000$ target degrees $\{k_1,k_2,\dots,k_N\}$ drawn randomly from a uniform distribution in $\{30,31,\dots,70\}$. Then we create links and triangles connecting nodes according to Eqs.~(\ref{eq37})-(\ref{eq40}) and generate a synthetic hypergraph. To each oscillator we assign a frequency drawn from a Lorentzian distribution with $\Delta = 1$ and $\omega_0 = 0$ by setting $\omega_n = \tan\left(\pi(2n - N -1)(N+1)\right)$.
For this hypergraph, we have $K_2^c \approx 0.038$ and $K_3^c \approx 0.021$.
\begin{figure}[t]
\includegraphics[clip,width=0.8\columnwidth]{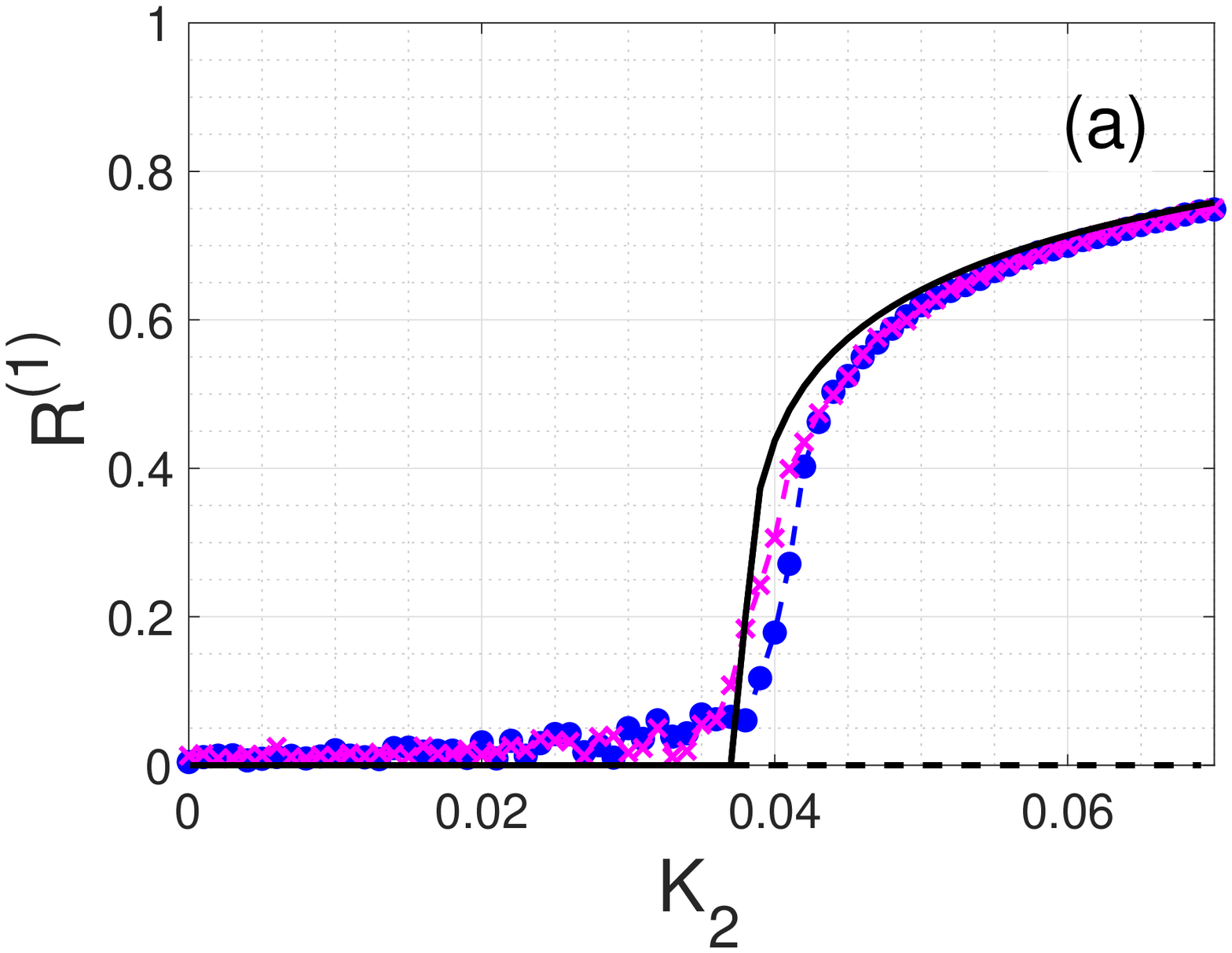}\\%
\includegraphics[clip,width=0.8\columnwidth]{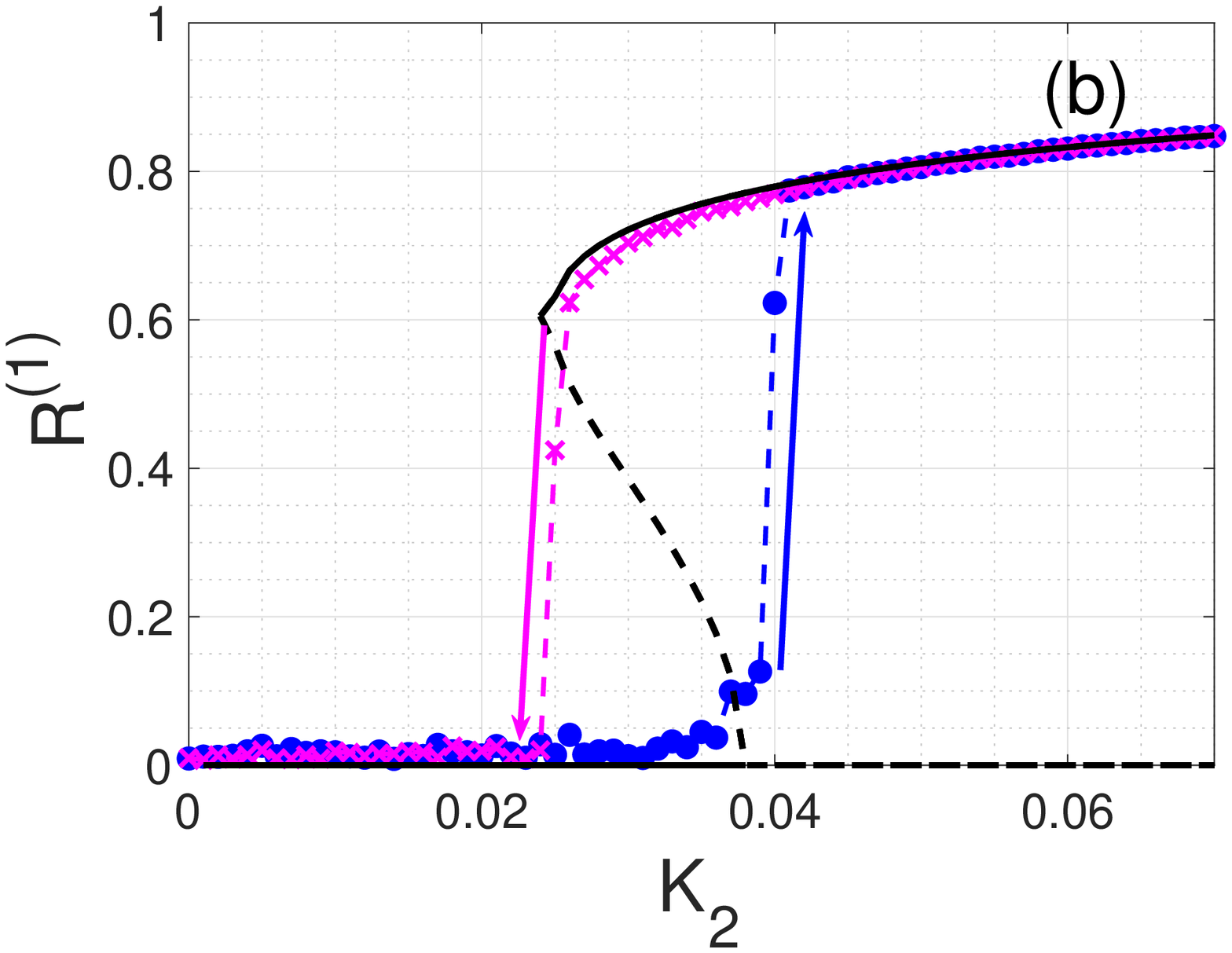}%
\caption{\label{fig1} The order parameter for a correlated hypergraph for two different cases: (a) $K_2< K_2^c$ and (b) $K_2 > K_2^c$. The numerical simulations are shown with the circled and crossed markers; the blue circles are the order parameter when gradually increasing $K_2$ whereas the pink crosses show the order parameter when gradually decreasing $K_2$. The solid black line is the stable order parameter and the dashed black line is the unstable order parameter found using mean field theory.}
\end{figure}

In Fig.~\ref{fig1} we show the steady-state value of $|R^{(1)}|$ as $K_2$ is adiabatically increased and then decreased (blue circles and pink crosses, respectively) for $K_3 = 0.02 < K_3^c$ [Fig.~\ref{fig1}(a)], and $K_3 = 0.05 > K_3^c$ [Fig.~\ref{fig1}(b)]. For each $K_2$, Eq.~\eqref{eq1} was solved numerically using Heun's method with a time step $\Delta t = 0.002$ for 100 time units, and the value of $|R^{(1)}|$ was averaged for the last 4 time units. 
For $K_3 = 0.02 < K_3^c$ [Fig.~\ref{fig1}(a)], the transition to synchronization is continuous and there is no hysteresis. On the other hand, for $K_3 = 0.05 > K_3^c$ [Fig.~\ref{fig1}(b)], the transition is explosive, and there is a hysteresis loop as $K_2$ is increased and then decreased (indicated with arrows). 
In general, the numerical solution of Eqs.~(\ref{eq1}) agrees well with the numerical solution of the self-consistent equations~(\ref{eq21})-(\ref{eq22}), shown as black lines, except for $K_2 \approx K_2^c$. The dashed black line corresponds to an unstable solution of Eqs.~(\ref{eq21})-(\ref{eq22}).
The observation that higher-order interactions promote bistability and hysteresis are consistent with findings in Refs.~\cite{skardal-arenas,skardal-arenas1}, where higher order interactions occur via all-to-all simplicial complexes. We note the discrepancy between the order parameters predicted by the mean-field theory and those calculated numerically. This could be a result of the finite network size we used for numerical simulations or our neglect of pair correlations in Eq.~(\ref{approximation}).

\begin{figure*}[ht]
\includegraphics[clip,width=0.7\columnwidth]{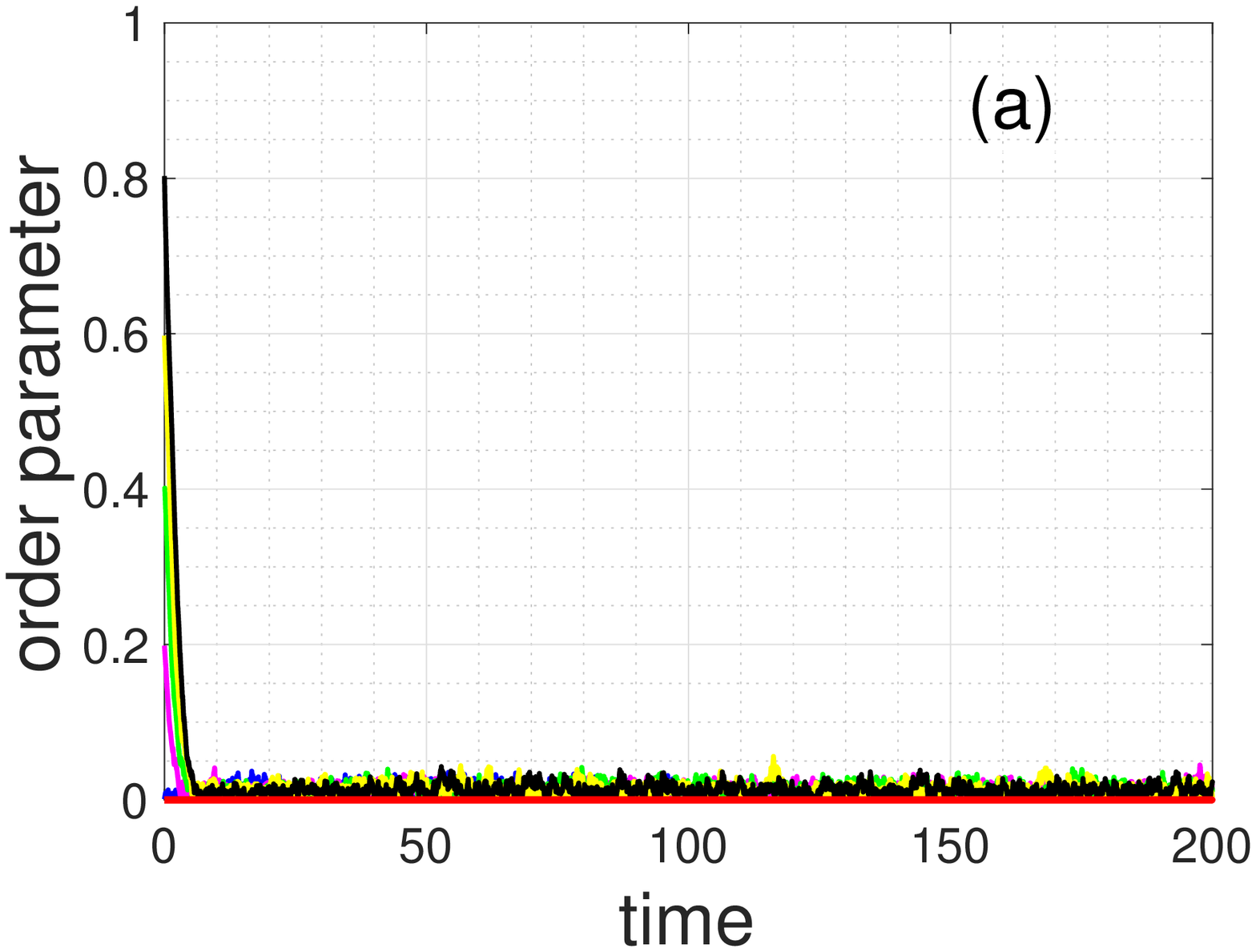}%
\includegraphics[clip,width=0.7\columnwidth]{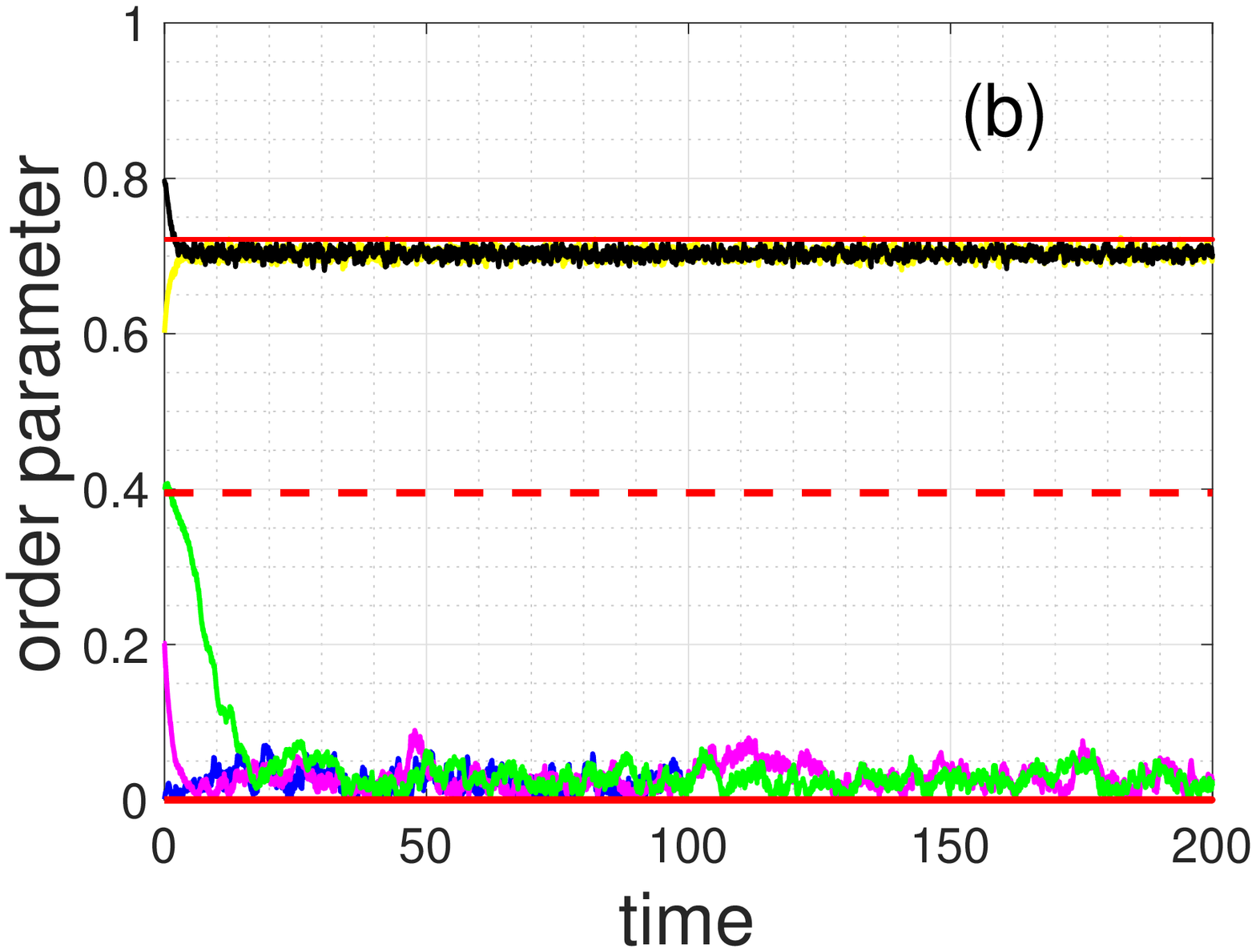}%
  \includegraphics[clip,width=0.7\columnwidth]{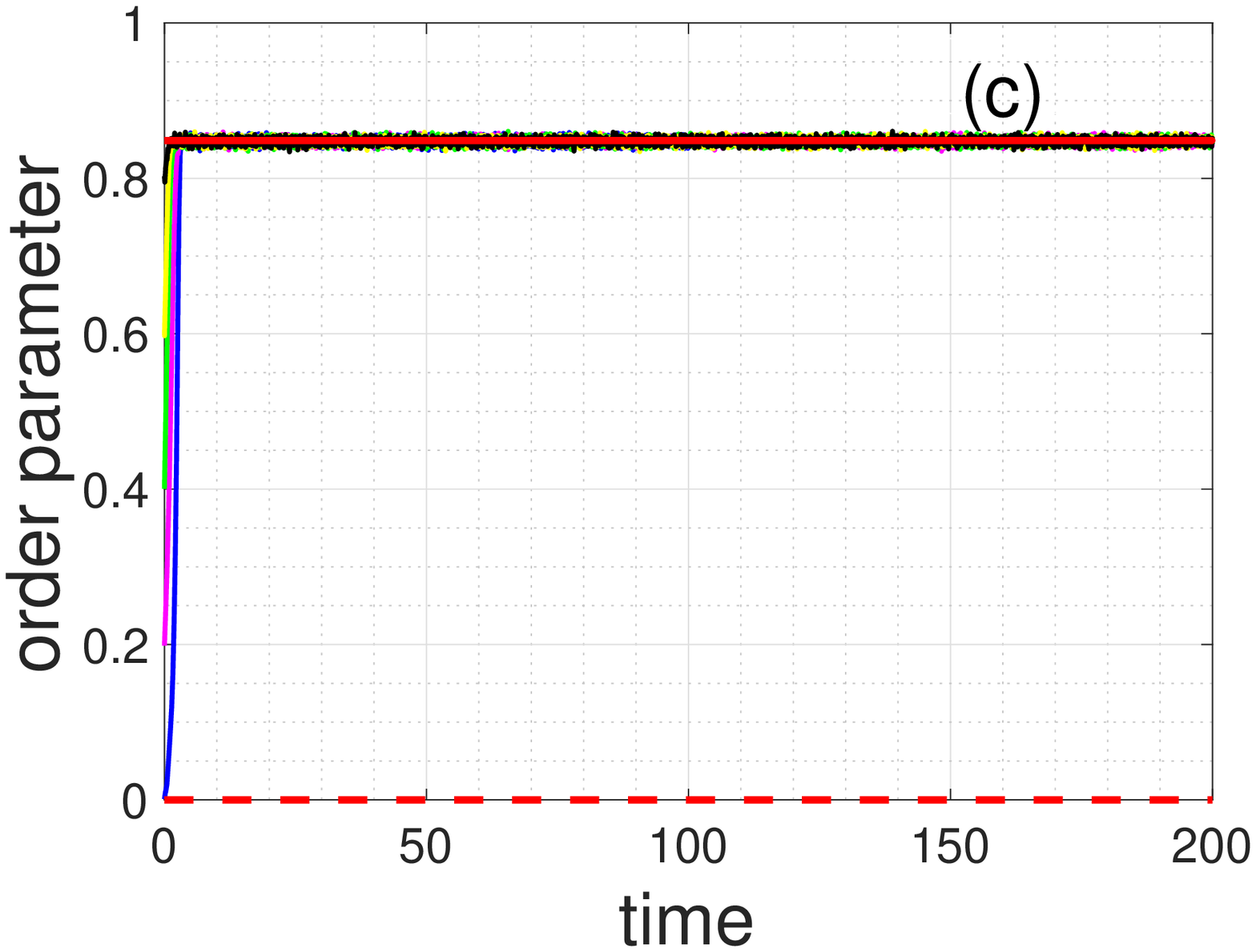}\\%
\caption{\label{fig3} The timeseries plot of the order parameter in three different regimes: (a) incoherent regime, (b) bistable regime, and (c) synchronized regime for a correlated hypergraph. The five different solid lines give the order parameter for five different initial conditions. The red solid line is the stable order parameter predicted by mean field theory whereas the dotted line is the unstable order parameter predicted by mean field theory.}
\end{figure*}

To further illustrate the bistable nature of the system, in Fig.~\ref{fig3} we plot $|R^{(1)}(t)|$ versus $t$ for fixed $K_3 = 0.05$ and $K_2 = 0.005$ (a), $K_2 = 0.03$ (b), and $K_2 = 0.07$ (c), corresponding to the incoherent, bistable, and synchronized regimes, respectively. For each value of $K_2$, we use five different initial conditions with $|R^{(1)}(0)| \approx 0, 0.2,0.4,0.6,0.8$. The solid (dashed) red lines indicate the stable (unstable) solutions of the steady-state self-consistent equations (\ref{eq21})-(\ref{eq22}). The values of $|R^{(1)}(t)|$ approach the values predicted by the mean-field theory, including both stable values in the bistable regime [Fig.~\ref{fig3}(b)].

\begin{figure}[b]
\includegraphics[clip,width=0.8\columnwidth]{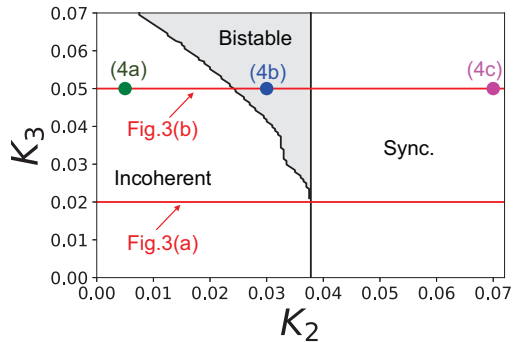}
\caption{\label{fig:newton} Phase diagram for the correlated hypergraph obtained from numerical solution of Eqs.~(\ref{eq20})- (\ref{eq22}). The vertical solid black line represents $K_2^c$. The lower horizontal solid red line corresponds to $K_3 = 0.02$ [Fig.~\ref{fig1}(a)]. The upper horizontal solid red line corresponds to $K_3 = 0.05$ [Fig.~\ref{fig1}(b)].}
\end{figure}

In Fig.~\ref{fig:newton} we present the phase diagram for this hypergraph model obtained from numerical solution of Eqs.~(\ref{eq20})-(\ref{eq22}). The horizontal red lines represent the parameters used in Figs.~\ref{fig1}, and the circles indicate the parameters used in Figs.~\ref{fig3}.

\section{Discussion} \label{Discussion}
In this paper we explored synchronization of phase oscillators on hypergraphs with heterogeneous structures, generalizing the results in \cite{skardal-arenas, skardal-arenas1} to more complex scenarios. The mean-field approximation  allowed us to predict the onset of synchronization, explosive transitions between synchronized and incoherent states, and their bistability as a function of system parameters. In the absence of hyperedges of size larger than 2, we recover a smooth transition between incoherent and synchronized states as found in the standard network Kuramoto model~\cite{Restrepo,rodrigues2016kuramoto}. Sufficiently strong higher order interactions lead to an abrupt transition and bistability of incoherent and synchronized states  (see Ref.\cite{kuehn2021universal} for a broader perspective of this issue). For a hypergraph with correlated links and triangles, we showed that the onset of synchronization and onset of bistability depend on the moments of the degree distribution. For the hypergraph model we considered, higher order-interactions only affect the onset of bistability, but not the onset of synchronization (however, see additional discussion on this point below). We have also verified that similar results hold true for networks with power law and bimodal degree distributions.

The main limitations of our study are the requirement for hypergraphs to be produced by the generative model of Sec.~\ref{Model}, the use of the mean-field approximation, and the use of approximation (\ref{approximation}). [Here, we refer to the approximation that all nodes with the same hyperdegree are statistically equivalent as the mean-field approximation, rather than neglecting pair correlations in Eq.~(\ref{approximation})]. The generative model we used assumes that the presence of a hyperedge connecting a group of nodes depends only on a set of pre-determined quantities of these nodes, which might not capture the generative mechanisms behind some real-world or model hypergraphs. For example, a simplicial complex model where triangles only join triples of nodes that are already forming a clique (as assumed in some studies \cite{iacopini2019simplicial,skardal-arenas1}) is not included in the class of models that the generative model in Sec.~\ref{Model} covers. In such a model, correlations between the states of nodes belonging to the same triangle could be non-negligible, and thus approximation (\ref{approximation}) could break down. In that case the techniques introduced in Ref.~\cite{hildebrand2007kinetic} could be needed to account for pair correlations. For example, for the SIS model on a simplicial complex, Ref.\cite{burgio2021network} finds that the epidemic threshold is only predicted correctly when accounting for pair correlations. In addition, Ref.~\cite{zhang2022higher} recently noted that synchronization properties in the strongly synchronized regime differ between simplicial complexes and random hypergraphs. Exploring the limitations and possible extensions of our method for simplicial complexes is an interesting problem left for future work.

Despite the limitations discussed above, our framework constitutes a flexible  method to study synchronization of phase oscillators on complex hypergraphs.  While we demonstrated our framework in a particular case [hypergraphs constructed following Eqs.~(\ref{eq37})-(\ref{eq40})], we emphasize that the techniques presented here allow for the study of a much larger class of systems. Examples include hypergraphs with independently chosen link and triangle degree distributions, correlations between hyperedge degrees and frequencies, and varying degrees of correlations between link and triangle degrees. The techniques presented here open a way to understand the effects that a large class of structural properties of hypergraph connectivity can have on the synchronization of coupled oscillators.

\acknowledgments

JGR and SA acknowledge useful discussions with Nicholas Landry. JGR and PSS acknowledge a useful discussion with Christian Bick. SA was partially supported by NSF grant DMS-2205967.

\nocite{*}
\bibliography{aipsamp}

\end{document}